\documentclass[aps,pre,twocolumn,superscriptaddress,letter,citeautoscript,nobalancelastpage]{revtex4}
\usepackage{graphicx}
\usepackage{times}

\newcommand{\mb}[1]{\mathbf{#1}}

\begin{document}
\title{Time-Dependent Transport Through Molecular Junctions}

\author{San-Huang Ke}
\affiliation{Department of Physics, Tongji University, Shanghai 200092, China}

\author{Rui Liu}
\affiliation{Department of Chemistry, Duke University, Durham, North Carolina
27708-0354, USA}

\author{Weitao Yang}
\affiliation{Department of Chemistry, Duke University, Durham, North Carolina
27708-0354, USA}

\author{Harold U. Baranger}
\affiliation{Department of Physics, Duke University, Durham, North Carolina
27708-0305, USA}

\begin{abstract}
We investigate transport properties of molecular junctions 
under two types of bias---a short time pulse or an AC bias---by combining a 
solution for the Green functions in the time domain with 
electronic structure information coming from \textit{ab initio}
density functional calculations. We find that the short time 
response depends on lead structure, bias voltage, and barrier heights
both at the molecule-lead contacts and within molecules. Under a low frequency 
AC bias, the electron flow either tracks or leads the
bias signal (capacitive or resistive response) 
depending on whether the junction is perfectly conducting or not. For high 
frequency, the current lags the bias signal due to the kinetic inductance. 
The transition frequency is an intrinsic property of the junctions. 
\end{abstract}
\maketitle

The goal of achieving the ultimate miniaturization of electronic components is the
driving force behind the realization of molecular electronic devices. 
The idea dates back to at least 1974 \cite{aviram}, and technology has advanced 
especially rapidly in the last decade 
\cite{Nitzan-Ratner-review,TaoReview2006}. While most nanoscale transport studies have focused on steady state behavior, recently the high frequency (GHz or THz) performance of nanotube and graphene diodes or transistors has been investigated
\cite{tube-frequency2,Manohara2005,Rosenblatt2005,JGuo2005,Plombon07063106,Gomet-Rojas072672,Chaste08525,Lin09422}.
The small junction areas, low capacitances, and high electron mobilities of these
molecular devices seem to offer a cutoff frequency in the $THz$ range \cite{Burke1981}.
From the theoretical point of view, the question of how molecules behave under
time-dependent perturbation has to be answered since the short time
response of functional units is essential to the construction of molecular
electronic devices. 

In recent years, different theoretical approaches have been developed for this purpose.  
The methods and schemes employed in these theoretical studies include, for example, time-dependent density
functional theory (TDDFT) combined with a Green function technique for open
model systems \cite{Kurth05035308}, a nonequilibrium Green function (NEGF) method
treating the time-dependent bias as a perturbation to the steady-state
Hamiltonian \cite{ZGuo2005}, real-time TDDFT propagation for closed
systems \cite{Cheng06155112,Sai07115410}, a quantum master equation scheme based on TDDFT for
model systems \cite{Li07075114}, a combination of TDDFT and NEGF with the wide-band limit
approximation for open systems \cite{Zheng07195127,Yam08495203}, real-time
propagation of the Kadanoff-Baym equations for open and
interacting model systems \cite{Myohanen0867001,Alexander08165112,Myohanen09115107}, 
and a self-consistent NEGF formalism 
for the electron transport through nanotubes under a time-dependent gate potential
\cite{Kienle09026601}.
Despite the large theoretical effort made by different research groups, the
computational studies for real open systems with atomic details described by 
{\it ab initio} electronic structure calculations
\cite{Zheng07195127,Yam08495203,Kienle09026601} are still lacking because of the
computational difficulty.
 
Within the Keldysh nonequilibrium Green function description \cite{Jauho1}, 
Jauho and coworkers have derived formulations of transport in the mesoscopic 
regime under influence of external time-dependent perturbations.  
Based on this work, Zhu and coworkers \cite{ZGuo2005} established a computationally efficient method without the need of the wide-band limit (WBL) approximation 
by using the zero bias equilibrium Green function as the initial state of a
tight-binding model system and carrying out the analysis in the time domain. The
finite correlation time in open systems reduces in large part the computational
effort and makes this method a practical approach for \textit{ab initio} study. 

In the present work, we follow this idea and extend the tight-binding level
theory to a density functional theory (DFT) description of the
electronic structure, i.e., the initial equilibrium states are obtained by 
a DFT-Green function formulation \cite{method3,method4,method5,method6,method1}. 
We investigate a relatively simple atomic chain system so that 
we can conveniently adjust various parameters in order to explore 
the general behavior of molecular junctions under a time-dependent bias, including 
a short time pulse and an AC bias. In this way, we discuss how the
time-dependent transport is affected by the
nature of the leads, the lead-molecule coupling, the barrier in the molecule, 
and the amplitude and frequency of the bias signal.

Following Ref. \cite{Jauho1,ZGuo2005}, the time dependent current is
\begin{equation}
\label{eqn:final}
I_{\alpha}(t)=\frac{2e}{\hbar}Re\mb{Tr}\int dt_{1}[G^{r}(t,t_{1})
\Sigma^{<}_{\alpha}(t_{1},t)+G^{<}(t,t_{1})\Sigma^{a}_{\alpha}(t_{1},t)],
\end{equation} 
where $\Sigma^{</a}(t_{1},t)$ is the lesser/advanced self-energy.
When no time-dependence is present, the steady state
Green function $\widetilde{G}$ is solved by using the NEGF technique in energy
space \cite{method1}. Under a time-dependent voltage, the single-particle
energies become time-dependent in the leads, which causes accumulation and
depletion of charges to form a dipole across the device region. 
Based on the known steady-state Green function $\widetilde{G}$, we solve for the 
retarded lesser Green functions by the Dyson and Keldysh equations, respectively \cite{Jauho1,ZGuo2005}. 

\begin{figure}[b] 
\includegraphics[width=3.0in,clip]{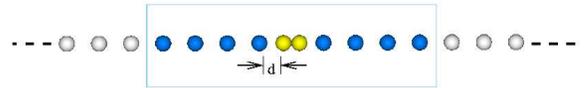} 
\caption{\label{fig:m2} Illustration of a diatomic hydrogen molecule (yellow) with $1\,$\AA\ bond length sandwiched between hydrogen chain leads. We vary the interatomic distance in the hydrogen chain but keep the distance between the H$_2$ and the leads fixed, $d\!=\!1.5\,$\AA. The H$_2$ molecule together with four hydrogen atoms on each side form the extended molecule (in the blue box). Bias is applied to the left lead.} 
\end{figure}

We first investigate the effect of lead structure on the transient response.
The system studied is a diatomic hydrogen molecule (H$_2$) with a $1\,$\AA\ bond
length sandwiched between two semi-infinite hydrogen chain leads as shown in 
Fig.~\ref{fig:m2}. The distance from the leads to the H$_2$ molecule is fixed to be 
$1.5\,$\AA. We change the H-H distance in the leads, thus varying the interatomic 
coupling strength. The narrow band width produced by a weak interatomic coupling 
constrains the electrons, producing less metallic leads. 

Single-zeta basis sets (SZ) and optimized Troullier-Martins pseudopotentials \cite{TM} 
are used for all calculations \cite{siesta}. For hydrogen, if the applied voltage 
is low so that \textit{p} states are not excited, a SZ basis set is a good approximation. The PBE version of the generalized gradient approximation (GGA) functional \cite{PBE} is adopted for exchange-correlation. For convenience, we adopt atomic units for electric current and time ($e$ = $\hbar$ = $m_e$ =1). An accurate description of atomic and chemical details in the contact region obtained from DFT calculations enables a meaningful study of realistic molecular devices.

\begin{figure}[t] 
\includegraphics[width=3.3in,clip]{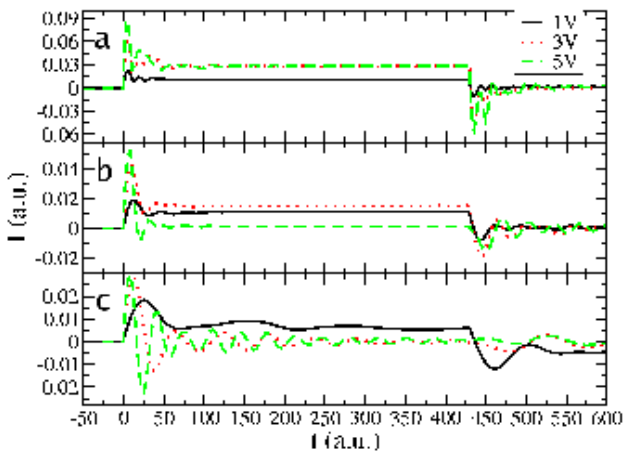} 
\caption{\label{fig:m2result} 
Current as a function of time (both in atomic units) for the H$_2$ molecule of
Fig.~\ref{fig:m2} and three different leads: the interatomic distance in the leads is (a) $1.5\,$\AA, (b) $2.0\,$\AA\, and (c) $2.5\,$\AA. A square shaped voltage pulse is applied to the left lead starting at $t\!=\!0$ and ending at $425$. Solid, dotted, and dashed lines correspond to applied voltages of $1$, $3$, and $5$ V, respectively.} 
\end{figure}

The $I(t)$ characteristics for leads with different H-H distances are shown 
in Fig.~\ref{fig:m2result}. A square shaped voltage pulse is applied to the left lead
starting at $t\!=\!0$ (a.u.) and ending at $t\!=\!425$. The time mesh varied between 200 and 600 points for all results shown here; generally, more points were needed for more strongly coupled leads.

The current initially increases for a short time and then oscillates because of interference for several to tens of a.u.\ depending on the type of leads. We call the time needed for the current to increase from zero to its maximum the finite current response time, $\tau_R$. This delay in response is related to the inertia (effective mass) of the carrier and provides a mechanism for inductance (kinetic inductance) different from the magnetic one. As $\tau_R$ is very short (of order 1 fs), the kinetic inductance is only observable at high frequencies.

The second feature to note in Fig.~\ref{fig:m2result} is that the initial rise in current is steeper when the interatomic distance in the leads is smaller 
(smaller $\tau_R$) for the same applied voltage. For instance, 1\,V bias, $\tau_R$ is about 4, 10, and 25 for interatomic distances 1.5, 2.0, and 2.5\,{\AA}, respectively. This behavior is easily understood by considering the effective mass of the electrons in these different leads: as the interatomic distance increases, the band width, of course, decreases, and so the electrons have a larger effective mass.   

\begin{figure}[t] 
\includegraphics[width=3.0in,clip]{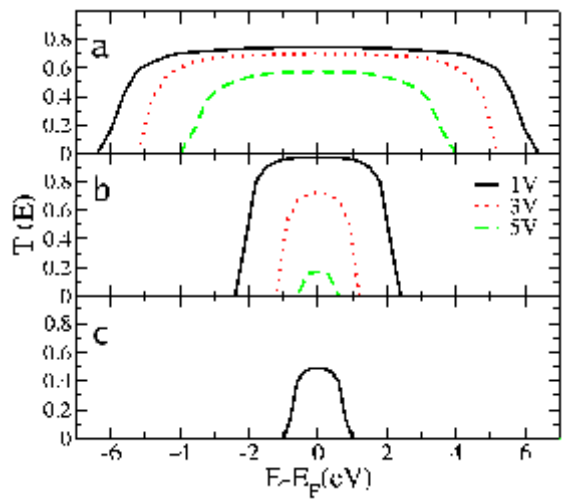} 
\caption{\label{fig:tf} 
Transmission as a function of energy for the H$_2$ molecule of Fig.~\ref{fig:m2} and the same cases as in Fig.~\ref{fig:m2result}. Solid, dotted, and dashed lines correspond to applied voltages of $1$, $3$, and $5$ V, respectively. 
}
\end{figure}

Fig.~\ref{fig:m2result} also shows the relations between bias voltage and the response time. For the leads with wider bands (band width wider than the bias window), $\tau_R$ is almost independent of the bias (panels a and b), indicating that the kinetic inductance is basically a constant. On the other hand, for the narrow band, less metallic lead (panel c), $\tau_R$ decreases rapidly for larger bias, showing that the kinetic inductance in this case depends strongly on the magnitude of the bias. One contributing factor is that the electrons are accelerating in the electric field; another is the mismatch between the narrow band in energy on left and that on the right when the bias window is larger than the band width (see Fig.~\ref{fig:tf} for further discussion). The light might occur, for instance, if a narrow band metallic oxide or silicide is used as the lead material. Over the duration of the transient response, the bias causes alternately accumulating and depleting regions of charge. Subsequently, the current finally tends to a steady state. 

In our system, a higher bias does not necessarily lead to a larger steady state
current---negative differential conductance can occur. In fact, the possibility
of highly nonlinear $I$-$V$ curves is one of the appealing features of molecular
electronics. Fig.~\ref{fig:tf} shows the transmission functions, $T(E)$, for the
same cases as Fig.~\ref{fig:m2result}. For reference, the widths of the s-band
in infinite hydrogen chains with interatomic distances $1.5\,$\AA, $2\,$\AA\,
and $2.5\,$\AA\ are $12.8\,$eV, $5.7\,$eV, and $2.4\,$eV, respectively. 
Focusing on the case with the narrowest band in the leads ($2.5\,$\AA\ interatomic separation), we see that the net current becomes nearly zero for a bias of $3\,$V (Fig.~\ref{fig:tf}c). This is because the s-bands for the left and right leads have no overlap when the bias is greater than the band width, $2.4$\,eV. Incident electrons encounter a hard wall, causing the current to oscillate over a much longer time than for wide bands. More generally, we see that the width of the transmission window equals the difference between the s-band width and the applied bias. 

\begin{figure}[t] 
\includegraphics[width=3.0in,clip]{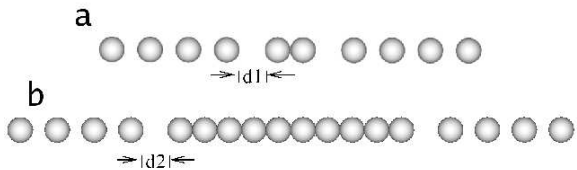} 
\caption{\label{fig:m15} Schematic of two extended-molecules: (a) H$_2$ and (b) H$_{10}$ shown with four atoms of each lead. $d1$ and $d2$ are the lead-molecule distances.} 
\end{figure}

The discussion in the last paragraph suggests the question: Does a less transparent junction generally lead to a longer time to reach a steady state? To answer this, we vary the barrier height associated with the molecule while keeping all other parameters constant. The barrier height at the molecule-lead contacts can be altered easily by changing the distance between the molecule and the leads. The barrier presented by the molecule itself can be changed by varying the number of atoms. Thus we shall compare results for H$_2$ and H$_{10}$ molecules (Fig.~\ref{fig:m15}).

For isolated H$_{2}$ and H$_{10}$ with an interatomic spacing of $1\,$\AA, the
gaps between the highest occupied molecular orbital (HOMO) and the lowest unoccupied molecular orbital (LUMO) are $10.7$\,eV and $3.2$\,eV, respectively (DFT GGA calculation). Fig.~\ref{fig:m15} shows only the extended molecules; note that the bias voltage is applied in the leads, i.e., outside of the extended-molecule region.

\begin{figure}[b] 
\includegraphics[width=3.3in,clip]{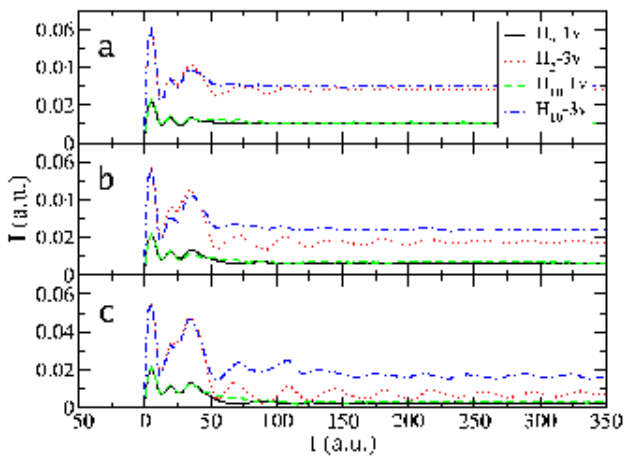} 
\caption{\label{fig:m1.5} Current as a function of time for both H$_{2}$ and H$_{10}$ molecules with different coupling strengths to the leads: the molecule-lead distance is (a) $1.5\,$\AA, (b) $1.7\,$\AA, and (c) $2\,$\AA. Results for a bias of both $1$\,V (H$_{2}$ solid, and H$_{10}$ dashed) and $3$\,V (H$_{2}$ dotted, and H$_{10}$ dot-dashed) are shown. The bias is applied at $t\!=\!0$; the interatomic distance in the leads is $1.5\,$\AA. 
}
\end{figure}

Fig.~\ref{fig:m1.5} shows the $I(t)$ curves of both molecules for three values of the lead-molecule separation. The bias voltage applied at $t\!=\!0$ is not turned off during the simulation. When the molecule-lead coupling is strong ($d1\!=\!d2\!=\!1.5\,$\AA, panel a), both the H$_{2}$ and H$_{10}$ junctions reach steady state quickly, but note that higher bias causes a longer transient regime. The magnitude of the steady state current is nearly the same for the H$_{2}$ and H$_{10}$ molecules.

\begin{figure}[b] 
\includegraphics[width=3.0in,clip]{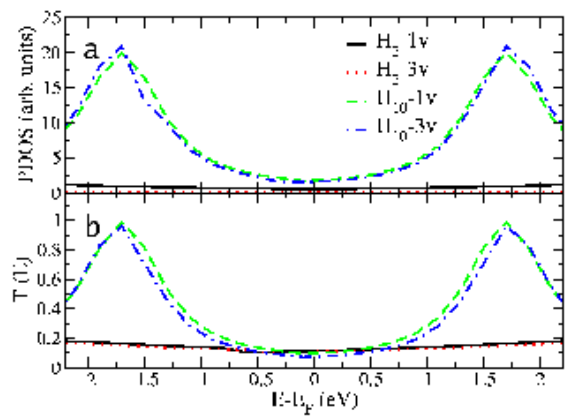} 
\caption{\label{fig:2ados} Projected density of states and transmission
functions for both the long and short molecules connected to hydrogen chain leads ($1.5\,$\AA\ interatomic distance). The molecule-lead distance is fixed at
$2\,$\AA. At energies more than 1\,eV away from $E_F$, the properties of H$_{2}$ molecular junctions ($1\,$V solid, $3\,$V dotted) differ greatly from those of H$_{10}$ ($1\,$V dashed, $3\,$V dot-dashed).}
\end{figure}

When the molecule-lead distance is stretched by $0.2\,$\AA\ to $1.7\,$\AA\ (panel b),
the low bias behavior of the molecules remains virtually the same, and further there is only modest quantitative change from the strongly coupled case. However, for a $3\,$V bias, the $I(t)$ curves differ substantially, both from each other and from the previous case. In the short molecule, the duration of the overshoot and oscillating regime is greatly extended, while in the long molecule, there is a surprisingly small decrease in the steady state current ($2.84 \!\times\! 10^{-2}$ in panel a to $2.25 \!\times\! 10^{-2}$ in b). H$_{2}$ experiences a more substantial change in current (from $2.70 \!\times\! 10^{-2}$ to $1.65 \!\times\! 10^{-2}$) in line with that expected from the fractional change in current at low bias. Thus, the higher molecule-lead barriers are more clearly manifest in the short molecule and hidden in the long one.

These differences are amplified further in the case of largest molecule-lead distance ($2\,$\AA, panel c). The low bias traces are quite similar to each other. At high bias, $I(t)$ for H$_{2}$ oscillates for a long time, while the steady state current for H$_{10}$ is surprisingly high.

We believe these difference between H$_{2}$ and H$_{10}$ are caused by the larger HOMO-LUMO gap in the short molecule. Fig.~\ref{fig:2ados} shows the projected density of states and transmission functions of both molecules in the weak coupling case (molecule-lead separation of $2\,$\AA). The $T(E)$ within the bias window for a $1\,$V bias (i.e. $-0.5\,$V to $0.5\,$V) are almost identical for H$_{2}$ and H$_{10}$. In contrast, for a $3\,$V bias, the $T(E)$ within the bias window, now $-1.5\,$V to $1.5\,$V, are totally different: for H$_{10}$, the tails of two resonant peaks extend into the bias window. Therefore the long molecule is much more transparent, leading to a larger current flow and less oscillatory behavior compared to the short molecule. Thus we see that both molecule-lead barriers and internal barriers within the molecules can cause significant changes in the $I(t)$ characteristics. 


\begin{figure}[t] 
\includegraphics[width=3.3in,clip]{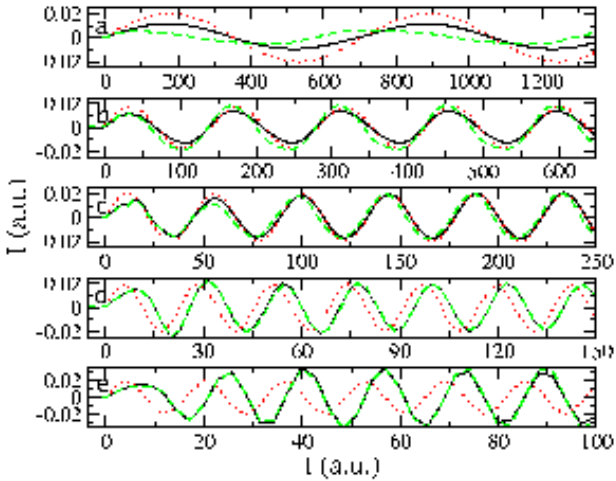} 
\caption{\label{fig:m15-peri} 
Current as a function of time for H$_{2}$ molecules subjected to AC signals of varying frequencies, smaller to higher from panel a to e. 
A well-coupled case (solid, $d1 \!=\! 1.5\,$\AA) and weakly-coupled case (dashed, 
$d1 \!=\! 2\,$\AA) are compared to the applied voltage (dotted, $1\,$V amplitude). The interatomic distance in the leads is $1.5\,$\AA.
}
\end{figure}

After studying the transient response to a square shaped pulse, we now look at 
how a molecular junction acts when a sinusoidal voltage with period $T$ is applied. 
The frequency should be slower than the plasma frequency $\omega_{P}$ of the
leads so that the electric field is effectively screened and the voltage drops
across the device region. For most metals, the plasma frequency is in the
ultraviolet regime, ranging from $10^{15}$ Hz to $10^{17}$ Hz, so typically 
this criterion is satisfied. For the 1D hydrogen chains in our calculations,
$\omega_{P}$ is of order $10^{17}$ Hz, assuming the dielectric constant and 
permeability are that of vacuum and the length of the device is around $1\,$nm \cite{Wang1989}. 

Fig.~\ref{fig:m15-peri} shows the AC response of H$_{2}$ junctions (schematic in 
Fig.~\ref{fig:m2}) with two molecule-lead distances, $1.5$ and $2\,$\AA, 
representing the well-coupled and weakly-coupled regime, respectively. First, 
consider the low frequency cases, $T/4 \!\gg\! \tau_R$ (panels a, b, and c). 
The current response of the well-coupled system tracks the sinusoidal signal, 
while for the weakly coupled junction, the current leads the bias voltage. 
This behavior can be understood by considering the equivalent electric circuit of the lead-molecule-lead system.
%
%
At low frequency, the inductance is not important, so we ignore it temporarily;
then we can view the junctions as a resistor ($R$) and capacitor ($C$) in parallel 
formed by the two contact interfaces. For the well-coupled system ($d1$ = 1.5 {\AA}), there is actually no interface, and thus the capacitor disappears, and the whole system is basically resistive. When the coupling is weaker, the capacitor is
formed, making the system capacitive in nature. 

As the frequency increases, the kinetic inductance gradually appears; consequently, 
the phase shift of the weakly-coupled system becomes progressively smaller in 
panels a, b, and c of Fig.~\ref{fig:m15-peri}. 
As the frequency increases further, $T/4$ becomes shorter, approaching the finite
current response time $\tau_R$. Then the effect of the delay in current becomes significant and at a certain frequency the characteristics of the junction 
changes from capacitive to inductive, as shown in panel d.
Our results are qualitatively consistent with a recent
calculation for an Al-nanotube-Al junction where it was found that the system is inductive for a high bias frequency ($T \sim$ response time) \cite{Yam08495203}. 
 

\begin{figure}[t] 
\includegraphics[width=3.3in,clip]{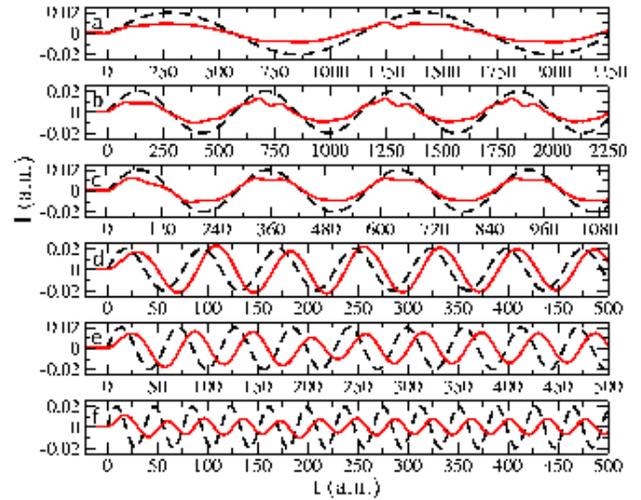} 
\caption{\label{fig:m25-peri} 
Current as a function of time (solid) for H$_{2}$ molecules subjected to AC signals (dashed) of varying frequencies, smaller to higher from panel a to f. 
The molecule-lead distance $d1$ equals $1.5\,$\AA, and the interatomic
distance in the leads is $2.5\,$\AA. 
}
\end{figure}

To support our explanation, we consider different values of $\tau_R$. Since we showed that $\tau_R$ in 1D molecular junctions is significantly affected by the nature of 
the leads, we consider a different lead structure but keep the intramolecular and lead-molecule structure the same. Thus the lead-molecule separation ($1.5\,$\AA) becomes different from the interatomic distance ($2.5\,$\AA) in the leads, 
and therefore a capacitor is formed.
The $I(t)$ curves in Fig.~\ref{fig:m25-peri} show that the current now leads the 
bias even for very small frequency (panels a and b). 
The current response changes, as expected, from capacitive to inductive
as the bias frequency increases; however, compared to the system with 1.5{\AA}
interatomic distance, the transition frequency decreases from $2.8 \times 10^{-2}$ 
(Fig.~\ref{fig:m15-peri}d) to $3.6 \times 10^{-3}$ (Fig.~\ref{fig:m25-peri}c). 
This reduction is consistent with the change in $\tau_R$: As the interatomic 
distance in the leads becomes larger, $\tau_R$ increases from $\sim\! 4$  to 
$\sim\! 25$ (Fig.~\ref{fig:m15-peri} a and c). 
As a caution, we mention that since the intrinsic magnetic inductance 
has not been taken into account here, the total inductance of the 1D molecular junctions may be underestimated; however, the magnetic inductance 
may be much smaller than the kinetic one for high frequencies (THz) 
\cite{Plombon07063106,Burke0355,Li04753,Yu051403,Rosenblatt05153111,Shang06668}.

It is interesting to note that the magnitude of the current driven by an AC bias
is not necessarily consistent with that driven by a DC bias. In
Fig.~\ref{fig:m1.5}, the current through H$_2$ with $d1\!=\!2.0$\AA\ is smaller than 
that with $d1\!=\!1.5$\AA. This is because the contact barrier in the former is higher. 
In contrast, under a high-frequency AC bias, the magnitudes of the current in 
these two cases become very close, as shown in Fig.~\ref{fig:m15-peri}.  
Although both bias frequencies are much slower than the plasma frequency $\omega_{P}$,
the time needed for the system to reach its steady
state is much longer, by orders of magnitude, than $1/\omega_{P}$ due to the
very weak screening in the 1-D structure. 
Only when $T/4$ of the AC bias is much longer than this time does the AC current
reflect the steady state current; otherwise, it will largely be determined 
by the transient states. 
Because the transient states are determined by the {\it extended molecule} instead of
the molecule and the contact alone, the magnitude of the AC current can
be very different from that of the DC current in the same system. 

\begin{figure}[t] 
\includegraphics[width=3.3in,clip]{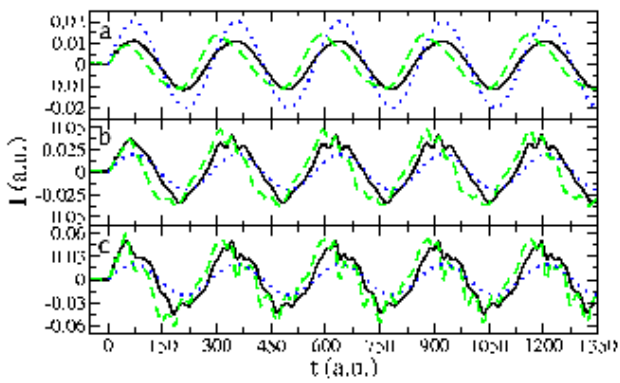} 
\caption{\label{fig:bias} 
Current as a function of time for H$_{2}$ molecules subjected to AC signals of different magnitude: 1\,V, 3\,V, and 5\,V for panels a, b, and c, respectively. 
$I(t)$ for two lead-molecule spacings are shown, $d1 \!=\! 1.5\,$\AA\ (solid) and $2\,$\AA\ (dashed), and compared to the applied voltage (dotted). 
The interatomic distance in the leads is $1.5\,$\AA.  
}
\end{figure}

Finally, we point out that when the AC frequency is relatively small, irregular 
behavior can be significant, as we saw for pulsed signals. We show in 
Fig.~\ref{fig:bias} that as the voltage magnitude increases (from 1\,V to 5\,V), the current response may not follow a sinusoidal wave. The irregular features are 
larger when the junction is less transparent. This is in line with our previous conclusion (Fig.~\ref{fig:m2result}c) and applies to the situation of relative 
low frequency AC response. 

In conclusion, the transient response of a molecular junction depends on the
lead structure, bias voltage, and barrier height seen by the transported electrons.
A higher electron density or a smaller effective mass leads to a faster response
characterized by a smaller current response time $\tau_R$. 
A high barrier height yields long oscillatory behavior in current, seen in
both the pulsed and AC situations. The current follows the AC
signal only when a junction is perfectly conducting and the signal frequency is
slow ($T/4 \!\gg\! \tau_R$); otherwise, a lead-molecule-lead junction  
should be viewed as a complex circuit consisting resistors, capacitors, and inductors.
Currents can lead or lag the AC signal, and the transition
frequency between the two regimes is an intrinsic property of the junction. 



\end{document}